\documentclass[conference]{IEEEtran}
\IEEEoverridecommandlockouts

\usepackage{cite}
\usepackage{amsmath,amssymb,amsfonts}
\usepackage{algorithmic}
\usepackage{graphicx}
\usepackage{textcomp}
\usepackage{xcolor}
\def\BibTeX{{\rm B\kern-.05em{\sc i\kern-.025em b}\kern-.08em
    T\kern-.1667em\lower.7ex\hbox{E}\kern-.125emX}}
\usepackage{import}
\usepackage{tabularx}
\usepackage{multirow}
\usepackage{booktabs}
\usepackage{array}
\usepackage{dblfloatfix}
\usepackage{listings}
\usepackage{url}
\usepackage{balance}
\definecolor{lstbg}{gray}{0.95}
\begin{document}

\title{Localisation and Circularity in Apple Supply Chains: An Algorithmic Exploration\\
}

\author{
\IEEEauthorblockN{Baraa Alabdulwahab }
\IEEEauthorblockA{
\textit{School of Computer Science}\\University of Bristol\\
Bristol, UK\\
fx25271@bristol.ac.uk
}
\and
\IEEEauthorblockN{Ruzanna Chitchyan}
\IEEEauthorblockA{
\textit{School of Computer Science}\\University of Bristol\\
Bristol, UK\\
ORCID 0000-0001-6293-3445
}
}

\maketitle

\begin{abstract}
Localisation and circularity in perishable food supply chains are essential for sustainability. Poor allocation of time-sensitive food leads to waste, higher transport emissions, and unnecessary long-distance sourcing. Algorithms used in digital trading platforms and allocation systems can help address these problems by improving how local supply is matched with demand under real operational constraints.

This paper examines localisation and circularity in the UK apple supply chain. Apples are an informative case because they are perishable, consumed fresh as dessert fruit, used as inputs across multiple food industries, and generate valuable by-products.

We present a weighted-sum mixed-integer linear programming formulation for supply--demand allocation. The model encodes a single global objective with explicit weights on four operational criteria: price matching, quantity alignment, freshness requirements, and geographic distance. These weights make priorities explicit and adjustable, enabling transparent balancing between economic and sustainability considerations. The framework also supports the circulation of unallocated supply across allocation cycles. Using a realistic apple supply--demand dataset, we evaluate allocation outcomes under different priority settings.

Results indicate that allocation outcomes are strongly shaped by both priority settings and the structure of the underlying supply network: explicit weighting enables controlled shifts in localisation, freshness alignment, and demand fulfilment, while revealing trade-offs between different priorities as they interact with the network’s unique characteristics.
\end{abstract}


\section{Introduction}\label{sec:intro}
Perishable food supply chains, including fresh fruit such as apples, involve products whose quality and economic value deteriorate over time. Allocation and distribution decisions are therefore time-sensitive and directly affect both sustainability outcomes (e.g., environmental impacts from underutilised produce and transport) and economic performance for sellers and buyers. Food losses and waste are widely recognised as a major sustainability challenge: the Food and Agriculture Organization of the United Nations (UN) estimates that roughly one-third of food produced for human consumption is lost or wasted globally, reflecting inefficiencies that occur across multiple stages of the supply chain \cite{FAO2011FoodLossWaste}. Beyond resource inefficiency, food that is not consumed also has a substantial climate footprint; the UN Environment Programme's Food Waste Index Report highlights that up to 10\% of global greenhouse-gas emissions is associated with food that is produced but not eaten \cite{UNEP2021FoodWasteIndex}. From a circular-economy (CE) perspective, distribution networks for perishables must therefore balance quality preservation, cost efficiency, and environmental performance. Presently, such distribution networks are increasingly operated through digital platforms, where allocation decisions are driven by resource-distribution algorithms. 

Sustainability-aware digital platforms and marketplaces require \emph{operational} decision support that can (i) encode circular-economy goals explicitly, (ii) remain computationally tractable, and (iii) make priority settings transparent when objectives change. While multi-objective optimisation approaches are widely used in sustainable food supply-chain design, many existing formulations treat circularity implicitly or offer limited transparency regarding how economic and environmental objectives are prioritised \cite{Azab2023,Belamkar2023}. Optimisation models that explicitly incorporate quality decay, such as first-expire-first-out (FEFO) strategies and shelf-life modelling \cite{Hertog2014ShelfLifeFEFO}, enable planners to coordinate production and distribution while managing freshness constraints. Mixed-integer formulations have been successfully used to integrate such quality considerations into supply-chain decision making \cite{Rong2011FreshFoodQualityMILP,Akkerman2010FoodDistributionReview}. 

In this paper, we build on this perspective by developing and evaluating a weighted-sum mixed integer linear programming (MILP) allocation model for the UK apple supply chain that operationalises circular-economy-relevant objectives. The proposed framework supports controlled balancing of economic performance, freshness requirements, and localisation (via geographic distance) within a single, deployable optimisation artefact suitable for repeated, short-term allocation decisions.

The key research questions addressed in this paper are:

\textbf{RQ1} how an optimisation framework can allocate UK-produced apples to demand while encoding circular-economy-relevant objectives alongside economic performance; and 

\textbf{RQ2} how different priority settings of these objectives influence allocation patterns, in particular localisation, supply utilisation, and distributional outcomes for buyers and growers.

The key findings from this study show that circular-economy-oriented allocation in
perishable food systems is fundamentally shaped by the structure of the underlying
supply network. Factors such as the geographic concentration of supply and demand,
pricing conventions, and seasonality-related attributes (e.g., freshness windows and
traded quantities) strongly influence which sustainability objectives can be jointly
achieved. As a result, no single static optimisation formulation can be considered
universally appropriate for circular food trading.

A second key finding is that, even with configurable and transparent priority settings,
a fully optimal and complete allocation is generally infeasible in operational trading
contexts. This arises from structural mismatches between supply and demand, as well as from the computational effort required to exhaustively search the allocation space.
From an ICT4S perspective, this implies that digital trading platforms should therefore
be designed around \emph{bounded, good-enough allocation} rather than exhaustive
optimisation, and should explicitly expose remaining incompatibilities through
diagnostic information. Furthermore, circular-economy-aware platforms should provide
systematic mechanisms to re-introduce unallocated quantities into subsequent trading
cycles, rather than treating non-allocation as unavoidable waste.

The remainder of this paper is structured as follows. Section~\ref{sec:RelatedWork} reviews related work on allocation mechanisms and circular-economy considerations in perishable food supply chains. Section~\ref{sec:problemDefinition} defines the allocation problem addressed in this study. The implementation of the proposed optimisation framework is described in Section~\ref{sec:implementation}, followed by its empirical evaluation under different priority settings in Section~\ref{sec:evaluation}. Section~\ref{sec:discussion} interprets the results with respect to their implications for resource distribution and circularity in the UK apple supply chain. Finally, Section~\ref{sec:conclusions} summarises the main contributions and outlines directions for future work.
\section{Related Work}\label{sec:RelatedWork}
Resource distribution and allocation algorithms for perishable food supply chains have been extensively studied, with mixed-integer linear programming (MILP) and multi-objective optimisation forming a dominant modelling paradigm. 
For instance, Azab et al.~\cite{Azab2023} propose a bi-objective MILP model for sustainable agro-food supply chain design, minimising total cost and carbon emissions while accounting for product perishability, seasonality, and waste disposal. The model integrates production, transportation, inventory, and waste management decisions and employs Pareto-front analysis to expose economic--environmental trade-offs. However, perishability is treated at a strategic level and does not explicitly address operational handling of downgraded products (e.g., reallocation from fresh consumption to lower-grade outlets).

Vanany et al.~\cite{Vanany2024} develop a multi-objective MILP model for a multi-echelon dairy supply chain, simultaneously minimising total cost, food waste, and environmental pollution. Perishability is incorporated through storage-time limitations and temperature-sensitive handling, and the model is evaluated using a real-world case study from Indonesia. While the results highlight the importance of cold-chain management, the analysis focuses primarily on strategic and tactical decisions, such as recommending delivery quantities and packaging configurations, rather than real-time operational coordination.

Liao et al.~\cite{Liao2025} present a multi-period, multi-objective optimisation model for perishable supply chains with quality decay, targeting profit maximisation and carbon emission minimisation. Using LP-metric and $\varepsilon$-constraint methods, the authors show that stricter emission constraints can substantially reduce environmental impact at the expense of economic performance. Food waste is treated indirectly via quality loss rather than as an explicit optimisation objective.

Mohammadi et al.~\cite{Mohammadi2024} propose a bi-objective nonlinear integer programming model for an IoT-enabled four-level perishable supply chain, minimising total cost and delivery time through joint optimisation of flows, quantities, and technology selection. Small instances are solved using GAMS, while larger ones are addressed with the Grey Wolf Optimizer (GWO). Results suggest that IoT adoption improves cost efficiency and responsiveness; however, the lack of validation on large-scale real-world data limits practical applicability.

Chiang et al.~\cite{Chiang2025} formulate a fuzzy multi-objective linear programming model for sustainable food delivery systems, balancing delivery cost, carbon emissions, delivery time, and food quality. Although focused on last-mile logistics, the study demonstrates how Pareto-based approaches can reduce emissions without compromising service quality. Perishability is treated implicitly through delivery-time constraints.

Kiani Mavi et al.~\cite{KianiMavi2025} introduce a multi-objective optimisation framework for sustainable food supply chain design that integrates economic, environmental, and social objectives with government policy instruments. Using NSGA-II, the author generate Pareto-optimal solutions illustrating policy impacts on cost and emissions. However, the model does not explicitly account for perishability or food quality decay.

Belamkar et al.~\cite{Belamkar2023} propose a multi-objective MILP framework for the design of the agro-food supply chain that focusses on trade-offs between economic performance and carbon emissions. Pareto-optimal solutions support strategic decision-making under competing objectives; however, perishability, freshness, and food waste dynamics are not explicitly modelled, limiting applicability to time-sensitive food products.

Across the reviewed studies, we observe two \textit{recurring limitations}: 

First, although sustainability and waste reduction are widely addressed, the explicit integration of circular economy mechanisms (such as recirculation of unallocated supply, quality downgrading, or rerouting expired products to alternative outlets) remains limited. Circularity is typically approximated through waste minimisation rather than operationalised as a continuous process.
Second, while multi-objective optimisation methods are common, most studies focus on Pareto-front generation or static weight configurations, typically at strategic or tactical planning levels, without examining how allocation outcomes respond to changing priorities in repeated operational settings. As a result, transparency regarding how economic and environmental priorities are enacted in operational decisions remains limited, reducing the ability of platform operators or policy-makers to adjust priorities under changing conditions.
\section{Defining the Problem}\label{sec:problemDefinition}

The core decision problem addressed in this paper is:
\emph{How should quantities of apples (or apple by-products) be allocated from available supply to demand in order to support circular-economy-relevant objectives (e.g., localisation and freshness management) while remaining economically viable for supply-chain participants?}

To model this problem in an abstract yet operational form, we consider an apple supply chain consisting of two sets of actors: sellers, who submit \textit{offers} of available product, and buyers, who submit \textit{orders} representing demand. Allocation decisions determine how quantities flow from individual offers to individual orders. Each potential allocation is characterised by multiple attributes, including allocated quantity, transaction price, geographic distance between seller and buyer, and temporal properties related to product freshness.

We therefore seek feasible flows from offers to orders subject to supply, demand, pricing, and logistics constraints, while allowing different priorities to be expressed over economic and sustainability-related criteria.

Two main algorithmic paradigms are commonly used for this type of allocation: (1) a weighted-sum (WS) formulation and (2) a Pareto-based (Pb) formulation. We present the Pareto-based approach for context and comparison, but focus on the weighted-sum formulation in the remainder of the paper.

\subsection{The Weighted-Sum Approach (Scalarised Optimisation)}\label{subsec:ws_approach}

In the WS approach \cite{BazganRuzika2022}, a weighted scoring function is defined over offer--order pairs based on individually scored matching attributes reflecting the operational criteria. The algorithm then seeks the optimal allocation of product flows that maximises the overall score:

\[
\text{score}(\text{offer}, \text{order}) 
= w_1 f_1(\cdot) + w_2 f_2(\cdot) + \dots + w_k f_k(\cdot)
\]

\noindent
where each \(f_k\) term captures a scored attribute, such as price deviation, shipping distance, or freshness alignment, and the weights \(w_k\) reflect the relative importance of these scores. The optimisation problem is then formulated as a single-objective assignment:

\[
\begin{aligned}
& \text{maximise} && f(x) = \sum_{i,j} \text{score}(\text{offer}_i, \text{order}_j) \, x_{ij} \\
& \text{subject to} && x_{ij} \in \mathcal{X},
\end{aligned}
\]

\noindent
where \(x_{ij}\) denotes the quantity allocated from \(\text{offer}_i\) to \(\text{order}_j\), and \(\mathcal{X}\) represents the set of feasible allocations defined by operational constraints.

The solution is optimal at the allocation \(x^* = \{x^*_{ij}\}\) that satisfies

\[
x^* = \arg\max_{x \in \mathcal{X}} f(x),
\]

\noindent
and the corresponding maximum objective value is given by

\[
f^* = f(x^*) = \sum_{i,j} \text{score}(\text{offer}_i, \text{order}_j) \, x^*_{ij}.
\]

\subsection{The Pareto-based Approach}

In the Pareto-based approach \cite{abir2020multi,Farag2020Hybrid,Nurjanni2017Green} the problem is formulated with multiple distinct objectives (for example: maximise total allocated quantities; minimise total buyer cost; maximise total seller revenue; minimise total shipping distance). Instead of converting into one aggregated objective, the Pb optimisation problem is solved by obtaining a Pareto‐optimal front of solutions (i.e., no solution can improve one objective without worsening another) .
\noindent
Assuming the decision variables \(x = \{x_{ij}\}\), representing allocated quantities from \(\text{offer}_i\) to \(\text{order}_j\), we have multiple objectives:
\[F(x) = (f_1(x), f_2(x), \dots, f_m(x))\]

\noindent
where each \(f_k(x)\) represents a different goal, for example:
\[
\begin{aligned}
f_1(x) &= \sum_{i,j} x_{ij} && \text{(total allocated quantity)}\\
f_2(x) &= - \sum_{i,j} \text{price}_{ij} \, x_{ij} && \text{(total buyer cost)}\\
f_3(x) &= \sum_{i,j} \text{revenue}_{ij} \, x_{ij} && \text{(total seller revenue)}\\
f_4(x) &= - \sum_{i,j} \text{distance}_{ij} \, x_{ij} && \text{(total shipping distance)}
\end{aligned}
\]

\noindent
The negative signs in $f_2(x)$ and $f_4(x)$ indicate that these objectives are minimization problems (rewritten as maximization problems for consistency). In the Pb approach, a solution $x^* \in \mathcal{X}$ is Pareto-optimal:


\[
\begin{aligned}
&x^* \text{ is Pareto-optimal if there is no } x \text{ such that} \\
&\quad f_i(x) \ge f_i(x^*) \ \forall i, \quad
f_j(x) > f_j(x^*) \ \text{for some } j .
\end{aligned}
\]

\noindent
The Pareto set is defined as

\[
\begin{aligned}
\mathcal{P} = \{\, x \in \mathcal{X} \mid\;
&\nexists\, y \in \mathcal{X} \text{ such that} \\
&f_k(y) \ge f_k(x)\ \forall k,\ \text{and} \\
&f_j(y) > f_j(x)\ \text{for some } j \,\}.
\end{aligned}
\]

Overall, both methods are valuable in supply-chain optimisation. Weighted-sum formulations are efficient and interpretable for daily operations, whereas multi-objective formulations are more suitable for long-term strategic planning where interactions between cost, quantity, and sustainability objectives must be explored systematically\cite{MarlerArora2010,Pappas2021}. 

\section{Defining Objectives}

We adopt a weighted-sum (WS) formulation rather than a Pareto-based approach because the objective of this study is \emph{repeated operational allocation} of perishable fruit (apples). The optimisation framework is intended to produce a single feasible allocation each time it runs, under time and data constraints, and the resulting decision must be straightforward to interpret  for circular economy stakeholders (including farmers, growers, juicers, etc..) and apply on an operational platform. This setting requires a method such as the WS formulation that yields one implementable solution, rather than a set of alternative solutions that would require additional selection or deliberation.

\subsection{Objective Structure}

As stated in the previous section, the weighted-sum formulation aggregates multiple allocation criteria into a single scalar score that represents the benefit of allocating quantity \(x_{ij}\) from offer \(i\) to order \(j\). Each criterion is represented by a normalised scoring function \(f_k(\text{offer}_i,\text{order}_j)\), scaled by a corresponding weight \(w_k\) that reflects its relative importance. These proxy functions are not intended to model detailed physical, behavioural, or biochemical processes. Instead, they provide simple representations of operational preferences (e.g., shorter distance, higher freshness, closer price alignment) that support fast, repeatable optimisation and transparent interpretation. The overall scoring function is defined as:
\[
\begin{aligned}
\text{score}(\text{offer}_i, \text{order}_j)
&= w_1 f_{\text{price}}
+ w_2 f_{\text{quantity}} \\
&\quad + w_3 f_{\text{freshness}}
+ w_4 f_{\text{distance}}
\end{aligned}
\]
\\
The individual components are defined as follows:

\begin{itemize}
    \item \textbf{Price alignment} (\(f_{\text{price}}\)): a normalised similarity measure between the offer per-unit price \(p_i\) and the order per-unit price \(p_j\). Higher scores are assigned when buyer and seller price expectations are closer, supporting feasible transactions and increasing the likelihood of repeated, stable trading. Exact matches receive the maximum score, while deviations are penalised proportionally; non-exact matches are allowed within feasible price bounds to reflect practical price tolerance, with buyer and seller minimum–maximum limits enforced as feasibility pre-conditions.
    \[
    f_{\text{price}} = 100 \times \left(1 - \frac{|p_i - p_j|}{\max(p_i,p_j)}\right)
    \]

    \item \textbf{Quantity alignment} (\(f_{\text{quantity}}\)): a similarity-based measure that rewards closer alignment between offered quantity \(q_i\) and requested quantity \(q_j\), while permitting partial matches. This promotes size-compatible matching (large quantities with large buyers and smaller quantities with smaller buyers), improving utilisation of heterogeneous supply and supporting participation by smaller growers, provided feasibility conditions (such as acceptable pricing) are met.
    \[
    f_{\text{quantity}} = 100 \times \left(1 - \frac{|q_i - q_j|}{\max(q_i,q_j)}\right)
    \]

    \item \textbf{Freshness alignment} (\(f_{\text{freshness}}\)): a linear measure of alignment defined by the difference, in days \(days_{ij}\), between the expiry dates specified in the offer and the order, normalised over one year. The expiry-date difference supports FEFO-style allocation by preferring offers that expire sooner, while still meeting buyer requirements; allocations are only considered if the offer’s expiry date is after the order’s required expiry date (a feasibility pre-condition applied before scoring). The linear formulation is used for simplicity and interpretability as a relative preference rather than a physical decay model. The one-year normalisation reflects typical upper bounds on apple storage under controlled conditions\cite{UMaineFruitColdStorage2026}.
    \[
    f_{\text{freshness}} = 100 \times \left(1 - \frac{days_{ij}}{365}\right)
    \]

    \item \textbf{Geographic distance} (\(f_{\text{distance}}\)): a distance-decay function that favours geographically proximate allocations. Distance is used as an operational proxy for transport effort and transport-related environmental impact. An exponential decay form reflects diminishing marginal preference for additional proximity: very short distances are strongly favoured, while differences between already long distances are less influential. The decay function is normalised using an upper-bound distance of 1400~km, corresponding to an approximate maximum road distance within the UK. This bound serves as a scaling parameter to ensure numerical stability and comparability across instances, rather than representing a physical routing constraint.
    \[
    f_{\text{distance}} = 100 \times \exp\!\left(-\frac{\ln(100)}{1400} \, d_{ij}\right)
    \]

\end{itemize} 

The resulting scalarised objective enables fast and interpretable optimisation, while making the effect of different priority settings explicit through the choice of weights \(w_k\).

\section{The Implementation}
\label{sec:implementation}

\subsection{Overview}
\label{sec:implementation_overview}

The implementation is structured into two main components:
\begin{enumerate}
    \item \textbf{Supply and demand entities}, represented as \texttt{Offer} and
    \texttt{Order} objects, encoding the market inputs and participant constraints.
    \item \textbf{An optimisation service}, which
    performs feasible flow construction, computes WS scores, solves the MILP allocation
    problem, and returns an allocation result (\texttt{Result}) with statistical analysis. It also handles unallocated offers and unmet orders, enabling systematic handling of residual quantities and circulation of unallocated supply across allocation iterations.

\end{enumerate}

The framework is implemented in Java, enabling portability across operating
systems and facilitating integration into web-based applications and
service-oriented architectures. The optimisation layer uses a MILP formulation
implemented with the \texttt{ojAlgo} optimisation library (v56.2.0), allowing the
framework to run without external solver dependencies.

\subsection{Supply and Demand Modelling}
\label{sec:implementation_supply_demand}

Supply and demand are modelled using two core domain entities: \texttt{Offer} (seller-side
supply) and \texttt{Order} (buyer-side demand). Both capture key constraints required for
perishable-food allocation, including quantity bounds, price bounds, time constraints,
and logistics metadata (postcodes). Additionally, each entity supports optional structural
preferences, such as whether a trade should be restricted to a single counterparty.

The offer (supply) and order (demand) structures are shown in Listing \ref{lst:offer} and Listing \ref{lst:demand} respectively.


\begin{lstlisting}[caption={Offer structure (simplified)},label={lst:offer}]
Offer:

  id, sellerId, productId
  
  quantity, minQuantity
  
  pricePerUnit, minPricePerUnit
  
  productionDate, expiryDate
  
  validFrom, validUntil
  
  logistics: collectionOnly, collectionPostcode
  
  flags: singleOrder
  
\end{lstlisting}


\begin{lstlisting}[caption={Order structure (simplified)},label={lst:demand}]
Order:

  id, buyerId, productId
  
  quantity, minQuantity
  
  pricePerUnit, maxPricePerUnit
  
  expiryDate, fulfillDate
  
  logistics: deliveryOnly, deliveryPostcode
  
  flags: singleOffer
  
\end{lstlisting}

These abstractions allow the framework to represent a broad range of supply-chain
participants (e.g., farms, storage facilities, wholesalers, markets) while keeping the
optimisation interface consistent: allocation is performed over feasible offer--order
pairs subject to the constraints encoded in these objects.

\subsection{The Optimisation Service Modelling}
\label{sec:implementation_optimisation_service}

The allocation process is implemented as a pipeline inside
\texttt{TradingOptimiserService.solve(...)}. Conceptually, the method consists of four
stages: feasible flow (arc) construction, WS scoring, MILP optimisation, and residual circulation.
Listing~\ref{lst:solve_pipeline} summarises the process.

\begin{lstlisting}[
  caption={Trading Optimisation Algorithm},
  label={lst:solve_pipeline}
  ]
 
Algorithm 1: TradingOptimiserService.solve(orders, offers, weights)

Input:
  orders: list of Order
  offers: list of Offer
  weights: [w_price, w_qty, w_expiry, w_dist], sum(weights)=1

Output:
  result: Result object containing allocated flows
  
Stage A: Feasible arc construction (match)
  A = empty set
  for each order o in orders:
    for each offer f in offers:
      if match(f, o) == true:
        add (f,o) to A
      else:
        record(f, o, reason)
        
Stage B: Scoring (Benefit creation)
  B = empty list
  for each (f,o) in A:
    s_price  = scorePrice(f, o)    // [0..100]
    s_qty    = scoreQuantity(f, o) // [0..100]
    s_expiry = scoreExpiry(f, o)   // [0..100]
    s_dist   = scoreDistance(f, o) // [0..100]
    score = w_price*s_price + w_qty*s_qty + w_expiry*s_expiry + w_dist*s_dist
    B.add( Benefit(f, o, score) )

Stage C: MILP optimisation (MILPSolver)
  result = MILPSolver.optimise(orders, offers, B)
  // MILPSolver implemented using ojAlgo v56.2.0
  // includes candidate pruning and LP-screening to reduce arcs

Stage D: Residual circulation
  subtract(result.flows, offers, orders)
  // offers become unallocated supply
  // orders become unfulfilled demand
  
  circulate (offers)
  // unallocated supply is assessed: carried forward if non-expired; otherwise flagged for downgrading/rerouting

  // reporting reasons and statistics
  
  return result
\end{lstlisting}
\paragraph{\textbf{Stage A: Feasible flow construction}}
The first stage constructs a feasible bipartite graph between offers and orders.
A candidate arc $(f,o)$ represents a potential flow, and it is included only if \texttt{match(f,o)} returns \texttt{true} (hard feasibility constraint).
This feasibility filter encodes domain constraints \emph{prior to optimisation}, ensuring
that the MILP considers only valid trades. Key feasibility checks include:
(i) product compatibility;
(ii) time-window feasibility (order fulfilment date within offer validity window);
(iii) expiry feasibility (offer expiry must be strictly after the order's required expiry,
supporting a minimum freshness buffer);
(iv) logistics feasibility (delivery-only vs.\ collection-only constraints);
(v) quantity feasibility, including minimum tradable quantities; and
(vi) price feasibility (overlap between seller and buyer acceptable price bounds).
If \texttt{match(f,o)} returns \texttt{false}, the reason is recorded for reporting purposes. 

\paragraph{\textbf{Stage B: Weighted-sum scoring}}
For each feasible pair, the framework computes four normalised scores in the range
$[0,100]$: price alignment, quantity similarity, expiry alignment, and distance
proximity. These are aggregated into a single score using the specified weights
(\texttt{weights} sum to 1). The resulting scalar score is stored in a \texttt{Benefit}
object and acts as the \emph{objective coefficient per unit} for that arc in the MILP.

\paragraph{\textbf{Stage C: MILP optimisation (MILPSolver)}}
The allocation is formulated as a mixed-integer linear program, implemented in
\texttt{MILPSolver}, and solved using \texttt{ojAlgo} library (v56.2.0). The main decision
variable is a continuous flow $x_{f,o}$ representing the allocated quantity from offer
$f$ to order $o$.

\noindent
The objective maximises the total weighted benefit:
\[
\max \sum_{(f,o)\in A} x_{f,o}\cdot \texttt{score}_{f,o}.
\]
The MILP enforces capacity and structural constraints, including:
(i) supply constraints per offer ($\sum_o x_{f,o} \leq \texttt{quantity}_f$),
(ii) demand constraints per order ($\sum_f x_{f,o} \leq \texttt{quantity}_o$),
(iii) minimum trade quantities (lower bounds on allocated quantities), and
(iv) optional single-counterparty constraints (\texttt{singleOffer}, \texttt{singleOrder}),
which ensure that an offer or order can be matched with at most one trading partner when set to true.

\textbf{Scalability enhancements:} In realistic datasets, the number of feasible arcs may approach $|\texttt{offers}|\times|\texttt{orders}|$. To ensure
tractable runtime, \texttt{MILPSolver} includes two practical enhancements:
\emph{candidate pruning} (restricting each order and offer to its top-$K$ scored matches),
and \emph{LP-screening} (solving a relaxed LP and discarding arcs with negligible relaxed
flow). These steps reduce the MILP size while preserving high-quality candidate matches.

\paragraph{\textbf{Stage D: Residual circulation and reporting}}
The \texttt{MILPSolver} output is returned as a \texttt{Result} containing selected flows (allocations) and summary information used for downstream evaluation. The framework compares the selected flows with the original offers and orders to identify residual (unallocated) quantities. For infeasible or unmet offer--order pairs, the recorded feasibility reasons and scores are reported, enabling participants to understand why trades failed or scored poorly and how to adjust their inputs. Non-expired unallocated supply is carried forward into a subsequent allocation iteration, supporting continuous circulation rather than disposal. By contrast, expired supply is flagged for quality downgrading and rerouting to alternative trading pathways (e.g., reclassifying edible but expired dessert apples as processing-grade apples for juicing, and subsequently to lower-grade outlets such as animal feed), thereby reducing avoidable waste.

The novelty of this implementation lies not in the use of MILP per se, but in integrating circular-economy-relevant priorities into an operational allocation pipeline. In particular, the framework combines: (i) explicit feasibility filtering prior to optimisation (e.g., expiry, price, and logistics constraints) with infeasibility reporting, (ii) transparent weighted-sum scoring that makes priority settings explicit and adjustable for platform operators, (iii) an MILP allocation formulation that executes these priorities as a single interpretable decision while preserving participant feasibility, and (iv) residual circulation that carries unallocated supply into subsequent allocation cycles.

All these elements support deployable, repeatable allocation for perishable food systems (i.e., the apple supply chain) and demonstrate how a trading platform can shift from purely economic matching to circularity-aware allocation.

\section{Evaluating the Framework}
\label{sec:evaluation}

\subsection{Evaluation setup}
\label{sec:evaluation_setup}

The evaluation of the proposed framework using the UK apples supply chain as a case study aims to assess whether (i) the optimiser produces allocations that satisfy all hard feasibility
constraints, and (ii) changing weighted-sum priorities produces \emph{directionally
consistent} changes in allocation outcomes (economic alignment, localisation, and
freshness alignment), while maintaining high domestic (i.e. UK-grown) supply utilisation.

\paragraph{Dataset preparation}
We derive the evaluation dataset from the ENG-Apple dataset (262 demand records and 2302
supply records; three apple types: edible, culinary, cider).\footnote{\url{https://github.com/horticulture-uk/ENG-Apple-dataset/}}
Each demand record is treated as an \texttt{Order}, each supply record as an \texttt{Offer},
and each apple type is treated as a separate product. We use the dataset quantities,
postcodes, and unit prices directly. Other inputs that are not provided by the dataset are generated as follows: (i) minimum tradable quantities are set to $0.01 \times$ the quantity;
(ii) offer minimum prices are random values between 80\%--90\% of unit price; and
(iii) order maximum prices are random values between 110\%--120\% of unit price. Order
expiry dates are random dates within $\pm 7$ days of 20 Oct 2025, with fulfilment dates set 7
days before expiry. Offer expiry dates are random dates within $\pm 7$ days of 27 Oct 2025,
with offer validity set to the full year. All IDs are generated as random unique
identifiers. All offers and orders accept partial quantities, and single-counterparty
flags are disabled.

Initial testing showed that solving the full multi-product instance (edible, culinary and cider apples) led to high MILP
runtime. Therefore, for this evaluation we restrict the instance to \textbf{edible apples} only, yielding \textbf{118 orders} and \textbf{932 offers}.

\paragraph{Iterative allocation protocol}
Each evaluation scenario runs the optimiser iteratively. As described in \emph{Optimisation Service Modelling} (Section \ref{sec:implementation_optimisation_service}), after each run, allocated quantities are subtracted from remaining offers and orders, and the next iteration resolves on the residual quantities. The process ends when no feasible matches remain or when the cap ($\texttt{maxIterations}=5$) is reached. In all reported scenarios, however, the solver terminated naturally after \textbf{3--4 iterations}.

\paragraph{Solver configuration.}
To limit the MILP size, we use candidate pruning with $\texttt{TopK}=250$.

\paragraph{Priority settings (weight vectors).}
We evaluate seven priority settings as `allocation strategies' (weights sum to 1):
\begin{itemize}
  \item Equal Weights: $(0.25,0.25,0.25,0.25)$
  \item Price First: $(\textbf{0.55},0.15,0.15,0.15)$
  \item Quantity First: $(0.15,\textbf{0.55},0.15,0.15)$
  \item Expiry First: $(0.15,0.15,\textbf{0.55},0.15)$
  \item Distance First: $(0.15,0.15,0.15,\textbf{0.55})$
  \item Price Extreme: $(\textbf{0.80},0.05,0.05,0.10)$
  \item Distance Extreme: $(0.10,0.05,0.05,\textbf{0.80})$
\end{itemize}

\paragraph{Reported metrics (interpretation)}
We report outcomes for both orders (buyers) and offers (sellers), focusing on the following metrics (Upward/downward arrows indicate whether higher/lower values are preferred for the allocation objective):
\begin{itemize}
    \item Supply utilisation (\(\uparrow\)), circulated supply (\(\uparrow\)), and leftover supply (\(\downarrow\));
    \item Distance (\(\downarrow\)): total distance required to ship the allocated quantity per order/offer;
    \item Average unit price: for orders (\(\downarrow\)), for offers (\(\uparrow\));
    \item Expiry-gap (\(\downarrow\)): prioritisation of near-expiry supply;
    \item SD of \emph{per-participant} allocation ratio (\(\downarrow\)): lower dispersion across participants indicates fairer allocation;
    \item Fragmentation (\(\downarrow\)): number of counterparts (flows) per order/offer.
\end{itemize}

\subsection{Results from allocation simulations}
\label{sec:evaluation_core_results}

Table~\ref{tab:eval_aggregate_new} summarises aggregate outcomes, while
Table~\ref{tab:eval_distribution_new} reports per-participant metrics (mean/SD/skewness)
for distance, average unit price, expiry-gap, allocation ratio, and fragmentation (number of counterparts).

\begin{table*}[t]
\centering
\footnotesize
\setlength{\tabcolsep}{3.0pt}
\renewcommand{\arraystretch}{1.10}
\begin{tabularx}{\textwidth}{>{\raggedright\arraybackslash}Xrrrrrrr}
\toprule
\textbf{Metric} &
\textbf{Eq} &
\textbf{Price} &
\textbf{Qty} &
\textbf{Expiry} &
\textbf{Dist} &
\textbf{PriceX} &
\textbf{DistX} \\
\midrule
Orders\# & 118 & 118 & 118 & 118 & 118 & 118 & 118 \\
Fully met orders\# & 47 & 48 & 52 & 49 & 29 & 37 & 24 \\
Partially met orders\# & 57 & 57 & 53 & 57 & 72 & 66 & 70 \\
Unmet orders\# & 14 & 13 & 13 & 12 & 17 & 15 & 24 \\
Offers\# & 932 & 932 & 932 & 932 & 932 & 932 & 932 \\
Fully met offers\# & 824 & 835 & 761 & 830 & 859 & 865 & 881 \\
Partially met offers\# & 16 & 16 & 23 & 17 & 12 & 9 & 8 \\
Unmet offers\# & 92 & 81 & 148 & 85 & 61 & 58 & 43 \\
\midrule
Total demand (tonnes) & 505198 & 505198 & 505198 & 505198 & 505198 & 505198 & 505198 \\
Total supply (tonnes) & 192768 & 192768 & 192768 & 192768 & 192768 & 192768 & 192768 \\
Total allocation (\textit{1st iteration}, tonnes) & 175818 & 177258 & 174869 & 175877 & 181616 & 179830 & 183044 \\
Supply utilisation (\textit{1st iteration}, \%) & 91.21 & 91.95 & 90.71 & 91.24 & 94.21 & 93.29 & 94.96 \\
Total allocation (\textit{at termination}, tonnes) & 184885 & 186068 & 184178 & 184793 & 185069 & 186531 & 186035 \\
Supply utilisation (\textit{at termination}, \%) & 95.91 & 96.52 & 95.54 & 95.86 & 96.01 & 96.76 & 96.51 \\
Circulated supply (\%) & 4.70 & 4.57 & 4.83 & 4.62 & 1.80 & 3.47 & 1.55 \\
Leftover supply (\%) & 4.09 & 3.48 & 4.46 & 4.14 & 3.99 & 3.24 & 3.49 \\
\bottomrule
\end{tabularx}
\caption{Aggregate evaluation outcomes across weighting strategies. (Eq = equal weights, Price/Qty/Expiry/Dist = first-priority, PriceX/DistX = extreme).}
\label{tab:eval_aggregate_new}
\end{table*}

\begin{table*}[t]
\centering
\footnotesize
\setlength{\tabcolsep}{4.0pt} 
\renewcommand{\arraystretch}{1.10}

\begin{tabularx}{\textwidth}{%
  >{\raggedright\arraybackslash}X
  @{\hspace{6pt}}r @{\hspace{6pt}}r @{\hspace{6pt}}r @{\hspace{6pt}}r
  @{\hspace{6pt}}r @{\hspace{6pt}}r @{\hspace{6pt}}r }

\toprule
\textbf{Metric} & \textbf{Eq} & \textbf{Price} & \textbf{Qty} & \textbf{Expiry} & \textbf{Dist} & \textbf{PriceX} & \textbf{DistX} \\
\midrule

\multicolumn{8}{l}{\textbf{Orders}}\\

\multirow{2}{*}{Counterpart (Mean)} 
 & \textbf{7.46} & \textbf{7.52} & \textbf{6.92} & \textbf{7.52} & \textbf{7.56} & \textbf{7.70} & \textbf{7.68} \\
 & \textit{8.32, 1.55} & \textit{7.74, 1.56} & \textit{9.26, 1.85} & \textit{7.98, 1.35} & \textit{7.90, 1.37} & \textit{6.98, 1.23} & \textit{8.28, 1.30} \\
\addlinespace[3pt]

\multirow{2}{*}{Allocation ratio (Mean)} 
 & \textbf{0.63} & \textbf{0.64} & \textbf{0.66} & \textbf{0.63} & \textbf{0.51} & \textbf{0.61} & \textbf{0.48} \\
 & \textit{0.43, -0.46} & \textit{0.42, -0.48} & \textit{0.43, -0.62} & \textit{0.42, -0.45} & \textit{0.43, 0.03} & \textit{0.41, -0.37} & \textit{0.43, 0.16} \\
\addlinespace[3pt]

\multirow{2}{*}{Unit price (£, Mean)} 
 & \textbf{941.35} & \textbf{949.40} & \textbf{949.79} & \textbf{959.24} & \textbf{912.22} & \textbf{931.03} & \textbf{847.89} \\
 & \textit{347.53, -2.30} & \textit{336.51, -2.42} & \textit{336.45, -2.43} & \textit{325.09, -2.56} & \textit{376.06, -1.99} & \textit{357.58, -2.18} & \textit{429.86, -1.45} \\
\addlinespace[3pt]

\multirow{2}{*}{Distance (km, Mean)} 
 & \textbf{864.70} & \textbf{887.71} & \textbf{950.39} & \textbf{940.58} & \textbf{678.85} & \textbf{857.41} & \textbf{669.43} \\
 & \textit{1473.74, 5.30} & \textit{1119.14, 2.38} & \textit{1600.94, 4.37} & \textit{1551.46, 4.34} & \textit{813.69, 1.76} & \textit{867.48, 1.51} & \textit{896.93, 2.47} \\
\addlinespace[3pt]

\multirow{2}{*}{Exp. days gap (day, Mean)} 
 & \textbf{5.75} & \textbf{5.95} & \textbf{5.97} & \textbf{5.57} & \textbf{5.69} & \textbf{5.92} & \textbf{5.58} \\
 & \textit{4.25, 0.46} & \textit{4.14, 0.41} & \textit{4.51, 0.45} & \textit{4.20, 0.69} & \textit{4.18, 0.39} & \textit{4.02, 0.29} & \textit{4.36, 0.49} \\

\midrule
\multicolumn{8}{l}{\textbf{Offers}}\\

\multirow{2}{*}{Counterpart (Mean)} 
 & \textbf{0.94} & \textbf{0.95} & \textbf{0.88} & \textbf{0.95} & \textbf{0.96} & \textbf{0.98} & \textbf{0.97} \\
 & \textit{0.38, -0.09} & \textit{0.35, -0.70} & \textit{0.43, -0.24} & \textit{0.37, -0.46} & \textit{0.29, -1.26} & \textit{0.32, -0.33} & \textit{0.26, 0.20} \\
\addlinespace[3pt]

\multirow{2}{*}{Allocation ratio (Mean)} 
 & \textbf{0.90} & \textbf{0.91} & \textbf{0.84} & \textbf{0.91} & \textbf{0.93} & \textbf{0.94} & \textbf{0.95} \\
 & \textit{0.30, -2.67} & \textit{0.28, -2.88} & \textit{0.37, -1.85} & \textit{0.29, -2.79} & \textit{0.25, -3.46} & \textit{0.24, -3.60} & \textit{0.21, -4.22} \\
\addlinespace[3pt]

\multirow{2}{*}{Unit price (£, Mean)} 
 & \textbf{954.73} & \textbf{968.95} & \textbf{892.20} & \textbf{962.64} & \textbf{986.90} & \textbf{995.79} & \textbf{1005.03} \\
 & \textit{318.00, -2.63} & \textit{301.32, -2.85} & \textit{389.29, -1.83} & \textit{307.16, -2.77} & \textit{263.76, -3.40} & \textit{259.54, -3.48} & \textit{224.12, -4.13} \\
\addlinespace[3pt]

\multirow{2}{*}{Distance (km, Mean)} 
 & \textbf{109.48} & \textbf{112.39} & \textbf{120.33} & \textbf{119.09} & \textbf{85.95} & \textbf{108.56} & \textbf{84.76} \\
 & \textit{129.46, 2.30} & \textit{120.32, 2.04} & \textit{145.25, 2.02} & \textit{136.46, 2.03} & \textit{99.71, 2.59} & \textit{107.87, 2.08} & \textit{97.63, 2.74} \\
\addlinespace[3pt]

\multirow{2}{*}{Exp. days gap (day, Mean)} 
 & \textbf{6.06} & \textbf{6.16} & \textbf{5.84} & \textbf{5.72} & \textbf{6.63} & \textbf{6.79} & \textbf{6.52} \\
 & \textit{4.89, 0.67} & \textit{4.86, 0.55} & \textit{5.09, 0.59} & \textit{4.67, 0.70} & \textit{5.04, 0.56} & \textit{5.03, 0.54} & \textit{4.88, 0.55} \\

\bottomrule
\end{tabularx}

\caption{Distributional metrics across strategies. Values in \textit{italics} represent SD and skewness respectively.}
\label{tab:eval_distribution_new}
\end{table*}

\subsection{Insights by strategy}
\label{sec:evaluation_insights_new}

\paragraph{\textbf{Equal Weights (baseline)}}
\emph{The equal-weights strategy yields a broadly balanced allocation in which most domestic apple supply is utilised, demand is partially met across many buyers, and no single dimension (price, distance, quantity, or freshness) dominates outcomes. It produces moderate fulfilment, moderate transport distances, and a fragmented matching structure that serves as a neutral reference point for comparison.}

The baseline allocates $184{,}885$ units (utilisation $95.91\%$), including $4.70\%$ of circulated supply, leaving $4.09\%$ of
apple supply unallocated. On the demand side, $47/118$ orders are fully met and
$14/118$ remain unmet, with a mean fulfilment ratio of $0.63$, indicating broadly
distributed partial fulfilment. Trading is fragmented at the order level with orders
matched to multiple suppliers on average (mean counterpart
count is $7.46$), while offers typically match to at most one
buyer. Mean order unit price is £$941.35$, and mean order total distance is $864.70$ km,
reflecting the accumulation of transport distance across fragmented matches. The mean
order expiry-gap of $5.75$ days provides the baseline freshness reference.

On the supply side, offers achieve a mean allocation ratio of $0.90$, indicating that
most supply is close to fully allocated, with residual supply concentrated in a minority
of offers. Mean offer total distance is $109.48$ km, and mean offer unit price is £$954.73$.

\paragraph{\textbf{Price First vs.\ Price Extreme}}
\emph{Price-focused strategies shift allocations toward economically favourable matches, improving seller-side price outcomes while maintaining high overall supply utilisation. Stronger price emphasis reshapes allocation patterns without materially improving localisation or freshness, and leads to slightly more fragmented sourcing on the buyer side.}

Both price-focused strategies preserve high utilisation (above $96\%$) and reduce
leftover supply relative to the baseline. The Price Extreme strategy utilises more supply in the first iteration with only $3.47\%$ allocated in subsequent cycles. On the supply side, mean offer unit price
increases monotonically as price weight increases (£$954.73 \rightarrow$ £$968.95 \rightarrow$ £$995.79$), indicating stronger economic outcomes
for sellers. Buyer-side price effects are non-monotonic: mean order prices rise under
Price First but decline under Price Extreme, reflecting changes in which offers dominate
allocation rather than uniform price pressure.

Distance outcomes remain close to baseline on both sides, and freshness alignment does
not materially improve under price emphasis. Fairness and fragmentation remain broadly
stable, with only marginal increases in the number of counterparts per order ($7.46 \rightarrow 7.52 \rightarrow 7.70$) under
stronger price prioritisation.

\paragraph{\textbf{Quantity First}}
\emph{The quantity-first strategy maximises demand-side fulfilment, increasing the number of buyers whose orders are fully or largely fulfilled. This comes at the cost of higher transport distances and reduced participation among growers, leaving a larger share of domestic supply unused and concentrating allocation among size-compatible offers.}

Quantity First achieves the strongest demand-side outcomes, with the highest number of
fully met orders ($52$ vs.\ $47$ baseline) and the highest mean fulfilment ratio ($0.66$ vs.\ $0.63$ baseline). This improvement is accompanied
by noticeable supply-side concentration: a larger number of offers remain unused ($92 \rightarrow 148$) and the
mean offer allocation ratio declines to $0.84$, indicating that allocation favours a subset of
size-compatible suppliers. Circulation of unallocated quantities, however, achieves the highest percentage ($4.83\%$) across all strategies proving its effectiveness to mitigate the bias in allocation.

Logistics burden increases under this strategy, with higher mean transport distances on
both sides. Prices shift asymmetrically: mean offer prices decrease (£$954.73 \rightarrow$ £$892.20$) while the mean order unit price stays high (£$949.79$), consistent with relaxing economic alignment to satisfy volume
compatibility. Freshness alignment does not improve, and although average fragmentation
is slightly reduced (order counterparts $7.46 \rightarrow
6.92$), variability increases (SD $8.32 \rightarrow 9.26$), indicating uneven matching patterns across
orders.

\paragraph{\textbf{Expiry First}}
\emph{Expiry-first allocation delivers the strongest freshness alignment by prioritising nearer-expiry apples, while largely preserving baseline levels of supply utilisation and buyer fulfilment. This improvement in FEFO-style allocation is achieved with limited impact on economic outcomes, but may require sourcing from more distant growers.}

Expiry First yields the lowest mean order expiry-gap ($5.57$ days), indicating the strongest FEFO-style
behaviour. This improvement is achieved without reducing throughput: supply utilisation
and leftover rates remain close to baseline, and demand-side fulfilment is largely
unchanged.

Transport distances increase on both sides, reflecting the need to source nearer-expiry
supply from more distant locations. Economic outcomes remain stable, with order and offer
prices close to baseline levels, and fragmentation patterns remain effectively unchanged.

\paragraph{\textbf{Distance First vs.\ Distance Extreme}}
\emph{Distance-focused strategies achieve the strongest localisation effects, minimising transport distances for both buyers and growers and increasing the share of domestic apples that find a local outlet. These gains are accompanied by thinner fulfilment across buyers, with supply distributed more evenly but in smaller quantities per order.}

Distance-focused strategies produce the clearest localisation effects, substantially
reducing mean transport distances on both sides ($109.48 \rightarrow 85.95 \rightarrow 84.76$ km for offers, and $864.70 \rightarrow 678.85 \rightarrow 669.43$ km for orders) and compressing the right tail of
long-distance outcomes (skew $5.30 \rightarrow 1.76$ and $2.47$). These strategies also improve supply-side utilisation, with fewer unmet offers ($92 \rightarrow 61 \rightarrow 43$) and higher mean allocation ratios ($0.90 \rightarrow 0.93 \rightarrow 0.95$). Circulation rates in these strategies are the lowest ($1.80\%$ and $1.55\%$) indicating that supply is mostly allocated within the first iteration.

The trade-off emerges on the demand side: fewer orders are fully met ($47 \rightarrow 29
\rightarrow 24$), and mean order fulfilment
ratios decline ($0.63 \rightarrow 0.51
\rightarrow 0.48$) as supply is spread across more buyers. Economic outcomes shift
systematically, with decreasing buyer-side prices (£$941.35 \rightarrow$ £$912.22 \rightarrow$ £$847.89$) and increasing seller-side prices (£$954.73 \rightarrow$ £$986.90 \rightarrow$ £$1005.03$) under
stronger localisation. Freshness alignment remains close to baseline, indicating that
distance prioritisation alone does not guarantee FEFO-style improvements.

\subsection{Summary}
\label{sec:evaluation_summary_new}

Across all seven strategies, the framework allocates a consistently high fraction of
domestic supply (utilisation $\approx 95.54\%$--$96.76\%$ including about 2--5\% of circulated supply) and terminates after 3--4
iterations. This indicates that repeated re-optimisation on residual quantities quickly
exhausts most feasible trading opportunities under the model constraints.

Changing the weight vector shifts outcomes in the expected qualitative directions:
Quantity First increases order fulfilment and fully met orders but leaves more offers
unused; Expiry First tightens FEFO-style alignment with near-baseline throughput; and
Distance-focused strategies reduce transport distances and increase the fraction of
domestic supply allocated, but distribute supply more thinly across buyers (more partial
fulfilment and fewer fully met orders). These results support the assumption that
the framework can operationalise circularity-relevant priorities (localisation and
circulation of unallocated domestic supply), and the evaluation shows how platform-level priority changes are
reflected in measurable allocation behaviour on the UK apples case study.
\section{Discussion}
\label{sec:discussion}

\subsection{Algorithmic modelling and design choices}

The primary aim of the proposed framework is to operationalise circular-economy-relevant priorities—such as localisation and freshness—within an allocation mechanism that remains economically viable for both buyers and sellers (the \textbf{RQ1}). Rather than optimising a single metric in isolation, the framework is designed to expose and control trade-offs between economic and circular-economy objectives under short-term operational constraints.

The optimisation model represents multiple objectives using a simplified weighted-sum formulation, where each objective is expressed as a similarity or alignment score between offers and orders. In reality, these objectives are \emph{not independent.} For example, price is often correlated with quantity through volume discounts, and distance is directly related to effective freshness through transport time. These interactions are not explicitly modelled in the framework. Instead, objectives are treated as separable components whose relative importance can be adjusted via weights.

This simplification is intentional and aligned with the scope of the paper. Accurately capturing all cross-objective interactions would require substantially more complex scoring functions and additional data. For the purpose of demonstrating how circular-economy objectives can be operationalised within an allocation algorithm, the adopted similarity-based scoring functions provide a suitable abstraction. They allow clear interpretation of optimisation behaviour while remaining computationally tractable and implementable as part of a decision-support tool.

From an implementation perspective, the framework exhibits four design features that are particularly relevant for \textit{operational} circular-economy decision support in digital trading platforms:

First, the allocation problem is solved as a \textit{single feasible allocation}, avoiding complex Pareto-front solutions. 

Second, the use of \textit{explicit and adjustable weights} ensures transparency and enables platform operators or policymakers to prioritise economic and circularity objectives. 

Third, the iterative allocation protocol enables \textit{circulation of leftover supply}, allowing partially matched offers to re-enter subsequent optimisation rounds rather than being discarded. 

Fourth, the optimisation service produces \textit{diagnostic information} that explains why certain offers and orders remain unmatched, which is essential for identifying actionable barriers to circularity (e.g., misaligned pricing or infeasible quantities).

\subsection{Interpretation of evaluation results}

\subsubsection{The validity of WS formulation} The evaluation results show that the framework produces directionally consistent allocation outcomes under different weight settings. When the weight of a given objective is increased, the corresponding alignment metric improves in the resulting allocations (e.g., lower transport distance under distance-first settings or tighter expiry gaps under expiry-first settings), without violating feasibility constraints. This confirms the validity of the scoring functions and their role in shaping allocation behaviour.

\subsubsection{Continuous allocation cycles}
The framework supports circulation of leftover quantities after each distribution round. Excluding distance-focused strategies, more than 50\% of leftover supply resulted from the first cycle was allocated in subsequent cycles, and the allocation process terminates naturally after three to four iterations in all strategies. These together indicate that most feasible matches are realised early and that remaining quantities are constrained by incompatibilities (e.g., non-overlapping price ranges between orders and offers). Such high utilisation and natural termination are important effectiveness indicators for short-term distribution decisions: the solver stops because feasible matches are exhausted, rather than due to externally imposed time or iteration limits.

\subsubsection{Strategy effectiveness and context dependence}
No single strategy or weight configuration can be identified as producing the most effective allocation across all metrics (the \textbf{RQ2}). For example, distance-focused strategies achieve higher domestic supply utilisation in some cases, but this effect is largely attributable to the geographic concentration of apple production and intake facilities in counties such as Kent, Herefordshire, and Worcestershire. In such contexts, prioritising distance naturally favours regions with dense supply, rather than reflecting a universally superior allocation principle. This reinforces the importance of transparent, adjustable weights, as allocation outcomes must be interpreted relative to the structure of the supply network.

\subsubsection{On price-focused matching}
Price-focused strategies were particularly effective at utilising available supply and producing successful matches. This behaviour can be explained by the strong alignment between demand and supply unit prices in the dataset. Demand prices have a mean of £1061.2 (SD 42.56, skew 0.079), while supply prices have a mean of £1062.84 (SD 43.57, skew 0.057). This near-uniformity results in high price compatibility across offers and orders in this dataset. Although this demonstrates the ability of the framework to exploit favourable economic targets, it also indicates that price-driven performance is highly dependent on the characteristics of the data set (or the supply network).

\subsubsection{Fairness considerations}
The framework exhibits acceptable fairness behaviour (i.e., balanced allocation ratios across participants) under most strategies. The average number of counterparts per order is approximately 7.5, closely aligning with the offer-to-order ratio in the dataset (932 offers to 118 orders, approximately 7.9). This suggests that allocations are generally distributed across multiple suppliers rather than being dominated by a small subset. However, this balance deteriorates under the quantity-first strategy, where concentrated supply and demand in specific regions lead the optimiser to favour matched-volume offer--order pairs. From a platform perspective, fairness is an important indicator that larger suppliers do not monopolise allocation outcomes, and that production from smaller growers can find viable routes within the trading network. 

\subsubsection{On freshness prioritisation}
The expiry-first strategy demonstrates behaviour consistent with a First-Expired–First-Out (FEFO) principle, prioritising offers with shorter remaining shelf-life. At the same time, other metrics remain close to the baseline. This can be plausibly attributed to the way the dataset was generated: the average expiry of offers was set approximately seven days ahead of the average expiry of orders. As a result, the mean expiry gap remains around six days across all strategies, with only a modest tightening under expiry prioritisation. This suggests that expiry-based prioritisation is most impactful when supply freshness is heterogeneous, such as when previously unallocated or returned products are reintroduced into circulation.

\subsubsection{Implications of dataset characteristics}
The ENG-Apple dataset used in the evaluation is synthetic, and its artificial uniformity and skewness influence the observed outcomes. Geographic concentration of supply, aligned price distributions, and controlled expiry offsets may have simplified the matching problem and shaped the effectiveness of specific strategies. While this enables controlled experimentation, transactional data that reflect the natural variability, seasonal effects, and structural imbalances of the UK apple supply chain would likely produce different allocation patterns under the same objective weights.

This confirms the previous note that allocation outcomes are shaped jointly by the optimisation formulation and the distinct characteristics of the supply network.

\subsection{Handling leftover supply}

Across all strategies, approximately 3--5\% of supply remains unallocated. Diagnostic outputs from the allocation algorithm indicate that these leftovers are primarily associated with poor price alignment between a subset of offers and orders (i.e., the acceptable price ranges of these offers and orders do not overlap). The mean of price-matching scores of this subset is below 20\%, while other scores have means above 50\%. 

Handling leftover supply can therefore be approached in a structured manner: as low matching scores and infeasibility reasons are explicitly identified and reported by the algorithm, allocation failures can be addressed by communicating these diagnostic signals to the relevant stakeholders. This enables constraints to be relaxed either manually or automatically (e.g., by widening acceptable price bounds when persistently low pricing scores are detected, or by re-routing expired products to alternative outlets). While such adaptive mechanisms are not fully implemented in this study, the proposed framework provides the necessary ground to support these extensions.

\section{Conclusions}
\label{sec:conclusions}

In this paper, we developed and evaluated an operational allocation framework for the UK apple supply chain that incorporates circular-economy-relevant priorities—such as localisation and freshness—alongside economic performance. The framework demonstrates how a weighted-sum optimisation formulation can be used for repeated allocation of perishable food supply, while making priority choices explicit and actionable for the trading platform operators.

A key insight from this modelling work is that allocation outcomes are jointly shaped by (i) the optimisation formulation and (ii) the structural characteristics of the underlying supply network. For example, the geographic concentration of apple supply and demand in the UK case directly influences outcomes under quantity- and distance-focused strategies. As a result, no single static set of priority settings under WS formulations can be considered optimal across contexts. Instead, circular-economy-oriented platforms require configurable and transparent allocation mechanisms that allow priorities to be adjusted in response to network structure and operational preferences or conditions.

Furthermore, the assessment reveals that fully optimal allocations are generally infeasible in realistic, large-scale operational settings. In this study, the introduction of candidate pruning and LP-screening techniques was necessary to manage the computational complexity of the real-world optimisation problem represented by the ENG-Apple dataset. This highlights the importance of designing decision-support tools around \emph{good-enough} solutions: producing interpretable allocations, exposing unresolved incompatibilities, and enabling systematic handling of unallocated quantities. Such an approach supports continuous circulation of resources in digital trading platforms, rather than treating computational limits as a justification for avoidable waste.

\balance
\bibliographystyle{IEEEtran}
\bibliography{OtherReferences}

\end{document}